\documentclass[12pt]{article}
\usepackage{epsfig}
\usepackage{graphicx}

\newcommand {\eqref} [1] {(\ref {#1})}

\newcommand{\beq}{\begin{equation}}   
\newcommand{\eeq}{\end{equation}}
\newcommand{\beqn}{\begin{eqnarray}}   
\newcommand{\eeqn}{\end{eqnarray}}

\newcommand {\slsh} [1] {\not{\hbox{\kern-2pt${#1}$}}}
\newcommand{\gsim}{\lower.7ex\hbox{$
\;\stackrel{\textstyle>}{\sim}\;$}}
\newcommand{\lsim}{\lower.7ex\hbox{$
\;\stackrel{\textstyle<}{\sim}\;$}}

\begin{document}

\begin{titlepage}

\begin{flushright}
FTPI-MINN-12/05, UMN-TH-3032/12\\
\end{flushright}

\vspace{0.7cm}

\begin{center}
{  \large \bf  A Chiral  SU(\boldmath{$N$}) Gauge Theory Planar Equivalent\\[1mm] to Super-Yang--Mills 
}
\end{center}
\vspace{0.6cm}

\begin{center}

 {\large A. Armoni\,$^\star$ and
    M. Shifman\,$^\diamond$}
\end {center}

\begin{center}

%\vspace{3mm}

{\it  
$^\star$\, Department of Physics, Swansea University
Singleton Park, Swansea, SA2 8PP, UK\\
$^\diamond$\,William I. Fine Theoretical Physics Institute, University of Minnesota,
Minneapolis, MN 55455, USA}\end{center}

\vspace{0.7cm}

\begin{center}
{\large\bf Abstract}
\end{center}

\hspace{0.3cm}
	We consider the dynamics of a strongly coupled SU$(N)$ chiral
gauge theory. By using its large-$N$ equivalence with ${\mathcal N} = 1$  super-Yang--Mills theory
we find the vacuum structure of the former. We also consider its
finite-$N$ dynamics.
	
\vspace{2cm}

\end{titlepage}

Till this day chiral theories continue to be one of poorly explored corners \cite{shi} of
Yang--Mills theories with massless spinors at strong coupling. 
The 't Hooft matching condition \cite{matching} and (qualitative) continuations from $R_3\times S_1 \to R_4$
\cite{munsal} are the only (and rather limited) tools available 
at the moment in theoretical analyses.
The simplest chiral theory has gauge group SU(2) 
and the fermion $\psi$ in the three-index symmetric representation (SU(2)-spin 3/2). This theory has no internal anomalies (nor global anomaly) and no Lorenz and gauge invariant 
mass term is possible \cite{sunsal}.

Another well-known example of a chiral theory is the SU(5) theory
with $k$ decuplets $\psi^{[ij]}$ and $k$ antiquintets $\chi_i$ of left-handed fermions.
Finally, one can mention the so-called {\em quiver} theories 
in which the gauge group is a product 
\beq
\mbox{SU$(N)_1\times$SU$(N)_2\times$....SU$(N)_k$}
\eeq
 and the set of the left-handed fermions consists of $k$ {\em bifundamentals} 
$$
\psi^{i_1}_{j_2}\,,\,\, \psi^{i_2}_{j_3}\,,\,\, ...\psi^{i_{k-1}}_{j_k}
\,,\,\, \psi^{i_{k}}_{j_1}\,.
$$
At $k=2$ the quiver theory is non-chiral, a 
gauge invariant mass term can be built. However, if $k\geq 3$ the quiver theory is chiral.
This theory is nothing other than an orbifold daughter of SU$(kN)$ minimal supersymmetric Yang--Mills theory \cite{munsal}.

In this paper we will consider an interesting example of a chiral theory
which so far escaped attention. This theory is a result of cross-breeding between two
orientifold daughters \cite{ASV} of ${\mathcal N} =1$ minimal supersymmetric Yang--Mills theory (also known as supersymmetric gluodynamics).
We will refer to it as hybrid. The hybrid theory per se is {\em not} orientifold daughter of anything.
The orientifold projection of operators
such as Tr$\lambda^2$ (where $\lambda$ is the gluino field) is
not defined in the hybrid theory.
 
 In studying the hybrid chiral theory
we will combine several ideas and methods relevant to nonperturbative QCD
and Yang--Mills theories with massless spinors at strong coupling in general, in 
addition to the planar equivalence between the minimal ${\mathcal N}=1$ supersymmetric Yang--Mills
and its orientifold daughters.

Consider a hybrid SU$(N$) chiral gauge theory with the following matter content: a
left-handed fermion  $\psi_{[ij]}$ transforming in the two-index {\em antisymmetric} representation
of the gauge group, a left-handed fermion $\chi^{\{ij\}}$ transforming in the
(conjugate) two-index {\em symmetric} representation of the gauge group and eight
left-handed fundamental fermions $\eta_i^A$
($A = 1,2,.., 8$), see Table 1.\footnote{This matter content is applicable at $N\geq 5$.
At $N=2$ antisymmetric fermions are color singlets; they decouple. Symmetric fermions
are equivalent to the adjoint representation, which is real. Hence,
the theory is self-consistent without introducing $\eta_i$'s  and is non-chiral. At $N=3$
 antisymmetric fermions are equivalent to antifundamental fermions. 
 Hence, the model to be considered has a symmetric field $\chi^{\{ij\}}$ and seven
 $\eta_i$'s. At $N=4$ the antisymmetric representation 
 $\psi_{[ij]}$ is in fact real, and can be discarded.}

This theory is obviously chiral since no gauge invariant fermion bilinears can be written.
It is self-consistent, i.e.
the gauge symmetry is
anomaly-free.
Indeed, the (internal) gauge anomaly is proportional to
\beq
\sum_R\left(
\sum_{\rm left} {\rm Tr}_R\, \left(T^a\left\{T^b\,,T^c\right\}\right)
-\sum_{\rm right} {\rm Tr}_R\, \left(T^a\left\{T^b\,,T^c\right\}\right)\right)
\label{chthone}
\label{2}
\eeq
where 
 $T^{a,b,c}$ denote the generators of the gauge group in the representation 
$R$ to which a given fermion belongs,
the sums run over all
 left-handed and right-handed fermions, respectively, and over all representations,
 and Tr$_R$ denotes the trace in the representation $R$.
Finally, 
the braces $\{...\}$ stand for the anticommutator. 
Note that if $T^a$ is the generator in the representation 
$R$, the generator  in the representation $\bar R$ is $-\tilde T^a$ 
where tilde means transposition.
In the theory we suggest for consideration, Eq.~(\ref{2})
reduces to
\beq
( N -4  ) - (N + 4) + 8 = 0\,.
\label{3}
\eeq

\begin{table}
\begin{center}
\begin{tabular}{|c|c|c|c|}
\hline
$\rule{0mm}{6mm}\psi_{[ij]}$ &  1 &1 &0\\[2mm] \hline
$\rule{0mm}{6mm}\chi^{\{ij\}}$ &$-1 $ &$1 $ & 0\\[2mm] \hline
$\rule{0mm}{6mm}\eta_i^A$ &$\frac{1}{2}$& $-\frac{N}{4}$& 1\\[2mm] \hline
\end{tabular}
\end{center}
\caption{\small The matter content of the chiral SU$(N)$ theory and its U(1) charges; the corresponding currents are defined in
(\ref{10}) and (\ref{11}).}
\end{table}

Let us first discuss the global symmetries of the model.
At  $N\to\infty$ the fundamental quarks are
unimportant. We will discuss them later on, and ignore them for the time being.
Then the theory has two U(1) symmetries, with the corresponding currents
\beq
j^{\dot\alpha\alpha}_{(\psi )} = \bar \psi^{\dot\alpha}\psi^\alpha\,,\qquad
j^{\dot\alpha\alpha}_{(\chi )} = \bar \chi^{\dot\alpha}\chi^\alpha\,.
\label{4}
\eeq
Each of the above currents is anomalous,
\beqn
\partial_{\alpha\dot\alpha} j^{\dot\alpha\alpha} &=&
 \partial_\mu\,j^\mu =
\left(\begin{array}{c}
N-2\,,\,\, \mbox{ for}\,\,\psi\\[2mm]
N+2\,,\,\, \mbox{ for}\,\,\chi\\[2mm]
\end{array}
\right)\times \frac{1}{32\pi^2} \, F_{\mu\nu}^a \tilde{F}^{\mu\nu\,a}\nonumber\\[3mm]
&\stackrel{N\to\infty}{\longrightarrow}& \frac{N}{32\pi^2} \, F_{\mu\nu}^a \tilde{F}^{\mu\nu\,a}
\eeqn
One can consider two linear combinations of the above currents
\beqn
j^{\dot\alpha\alpha}_{(1)}
&=&
j^{\dot\alpha\alpha}_{(\psi )} - 
%\left(1-\frac{4}{N+2}\right)
j^{\dot\alpha\alpha}_{(\chi )},,
\nonumber\\[2mm]
j^{\dot\alpha\alpha}_{(2)}
&=&
j^{\dot\alpha\alpha}_{(\psi )} + j^{\dot\alpha\alpha}_{(\chi )}\,.
\label{6}
\eeqn
The first current is anomaly-free at $N=\infty$ , while the second is anomalous,
\beq
\partial_{\alpha\dot\alpha}\,  j^{\dot\alpha\alpha}_{(2)} = \, \,\frac{N}{16\pi^2} \, F_{\mu\nu}^a \tilde{F}^{\mu\nu\,a}\,.
\eeq
The current $j^{\dot\alpha\alpha}_{(1)}$ plays the role of a vector current, while 
$j^{\dot\alpha\alpha}_{(2)}$  plays the role of an axial current. The remnant of the latter is the discrete $Z_{2N}$
symmetry, which, as we will argue below, is broken down to $Z_2$ presumably by the condensate
$\left\langle \chi^{\{ij\}} F_j^k \psi_{[ik]}\right\rangle \neq 0$.

We wish to argue that the planar hybrid theory is equivalent,
 in a well-defined glueball sector, to planar ${\mathcal N}=1$ super-Yang--Mills.
  The equivalence of an SU$(N)$ theory with a single Dirac fermion in the two-index antisymmetric representation 
  (or a theory with a fermion in the symmetric representation) with ${\mathcal N}=1$ super Yang-Mills was 
 demonstrated  in \cite{ASV}. 
 
\begin{figure}[!ht]
\centerline{\includegraphics[width=8cm]{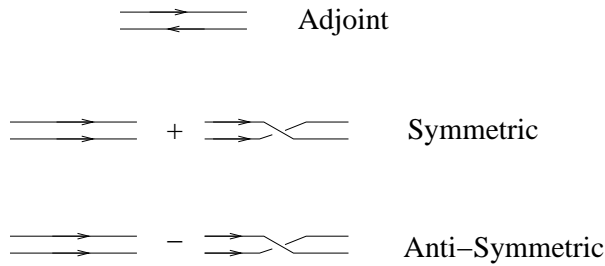}}
\caption{\footnotesize The 't Hooft double index notation for the fermion propagator    in either the adjoint, symmetric or  antisymmetric representations in a U$(N)$ gauge theory.} \label{feynman}
\end{figure}

 The reason for the perturbative equivalence is easy to understand: there is a one-to-one correspondence between the planar graphs of the two theories \cite{ASV}. Moreover, at the planar diagrammatic level there is no difference between symmetric fermions or antisymmetric fermions. The difference between the two representations arises 
 when fermion lines (in the the 't~Hooft double-index notation) cross, see Fig.~\ref{feynman}. These lines, however, do not 
 cross in any planar graph. 
 For this reason the hybrid theory is  perturbatively planar equivalent to ${\mathcal N}=1$ super-Yang--Mills.
 
The necessary and sufficient condition for {\it nonperturbative} planar equivalence 
between our hybrid theory and the minimal ${\mathcal N}=1$ super-Yang--Mills 
is the charge conjugation invariance ($C$-invariance) of the vacuum states \cite{unsal2}.
Note that the hybrid theory at finite $N$ is {\em not} $C$ invariant. However, $C$-invariance is 
restored at $N=\infty$. In the $N=\infty$  limit the nonperturbative dynamics of two theories --
the first with the Dirac fermion in the two-index symmetric representation and the second
with the Dirac fermion in the two-index antisymmetric representation -- are identical because the two representations -- symmetric and  antisymmetric -- become the same representation. Indeed, all the Casimirs 
coefficients of the two representations coincide in the limit $N\rightarrow \infty$. As a result, their dynamics are
 identical to the dynamics of the hybrid theory we consider here. The dynamics of the  three theories above
 become equivalent (in the  sector of glueball operators) to the dynamics of a vector theory with one Majorana
fermion in the adjoint representation (i.e. the minimal ${\mathcal N}=1$ super-Yang--Mills).

The implications of the exact  planar equivalence between the hybrid theory and ${\mathcal N}=1$ 
supersymmetric gluodynamics is the coincidence of the vacuum structure as well as the bosonic glueball
 spectra and dynamics in these  theories.  
The parent ${\mathcal N}=1$ theory has $N$ discrete vacuum states (see e.g. \cite{shi}), corresponding to the
breaking $Z_{2N}\to Z_2$ and labeled by the order parameter $\langle \lambda^2\rangle\neq 0$.
The same vacuum structure should be valid in the hybrid theory at $N=\infty$. An order parameter for the breaking is $\left\langle \chi^{\{ij\}} F_j^k \psi_{[ik]}\right\rangle$. The height of the ``barriers" separating these 
vacua is expected to be $O(N^2)$ \cite{GS}.

 The reason for coincidence of the bosonic glueball spectra is as follows. Let us integrate over the fermions of the hybrid theory. The resulting partition function is 
\beq
{\cal Z}= \int DA_\mu \exp( -S_{\rm YM}) \prod _{R,f} \left ( \det( \slsh \!{\mathcal D}_{\rm R}) \right ) ^{\frac{1}{2}} \, , \label{det}
\eeq 
where the above partition function \eqref{det} contains a product of determinants over the representations and flavors in the theory. In the planar limit the partition function \eqref{det} of the hybrid theory coincides with the partition functions of ${\cal N}=1$ super-Yang--Mills and the vector-like orientifold theories. For this reason all two-point functions of the form
\beq
\langle {\rm Tr}\,F^2 (x) ,  {\rm Tr}\,F^2 (y) \rangle \,, \, \,\,\, \langle {\rm Tr}\,F \tilde F (x) ,  {\rm Tr}\,F \tilde F (y) \rangle\,,\quad \langle {\rm Tr}\,F^3 (x) ,  {\rm Tr}\,F^3 (y) \rangle\,,
\eeq 
and so on, coincide in all four theories;  hence so do the glueball spectra. The only caveat in the above procedure is that in the hybrid theory it is impossible to introduce an infrared cut-off in the form of a mass term. That should not be a problem since the physical infrared spectrum is expected to develop a mass gap. Moreover, the parity degeneracies in the glueball spectra noted \cite{MSH} in supersymmetric gluodynamics and its orientifold daughters are inherited by the hybrid theory too.

\vspace{2mm}
  
Now, let us switch on $1/N$ corrections\,\footnote{A related discussion of possible phases of 
the chiral gauge theories can be found in \cite{Appelquist:2000qg}.} 
and address the most intriguing question of the chiral symmetry
implementation in the sector of 8 fundamental
fermion fields $\eta_i^A$.
The global symmetry of this sector of our hybrid theory is obviously
SU(8), in addition to a U(1) symmetry which we will   consider shortly.
No local color invariant bosonic operator containing two $\eta$ fields
(without $\bar\eta$'s)
 and an arbitrary number of other operators
exists. It is tempting to conclude that the chiral SU(8) is not spontaneously broken.

This conclusion is not likely to materialize, however. First, it goes against a (qualitative) argument 
due to Casher \cite{Casher:1979vw} that in strong coupling Yang--Mills theories with massless quarks 
confinement is impossible unless the chiral symmetry is spontaneously broken\,\footnote{Supersymmetric 
theories with confinement and no spontaneous breaking of a chiral 
symmetry are known, but this is because of the
presence of scalar quark fields which obviously negate the Casher argument.}
 (for a review see e.g. \cite{shi}). Second, if the chiral symmetry is unbroken, the 't Hooft 
 matching must be realized through
 saturation of the anomalous triangles by massless composite-fermion loops. A simple  reflection shows
 that there is no way to achieve such a saturation\,\footnote{
 This is despite the fact that,
unlike QCD, in the hybrid theory, even at large $N$, there exist three-quark spin-1/2
baryons, for instance,
$
\eta_{i\,,\beta}^{\{A} \,\eta_j^{B\}\,\beta}\, \chi^{\{ij\}}_\alpha\,,\quad
\eta_{i\,,\beta}^{[A} \,\eta^{B]}_{j\,\alpha}\, \chi^{\{ij\}\,\beta}
$. 
The $N$ factors still do not match in the comparison of 
the ``quark" and ``hadron" triangles. Warning: in the literature one can find reasonable arguments
 \cite{AR}
 against the ``straightforward" saturation. }
  at large $N$. 
 
In view of the above, let us examine less trivial operators for the role of order parameters for the
SU(8) chiral symmetry breaking. 

Using $\eta $ and $\, \bar \eta$ one can build, in principle,
a Lorentz and gauge invariant  order parameter whose expectation value could break SU(8),
for instance,
\beq
{\mathcal O}^A_B= \eta^{\alpha\, A}_i\, {\bar \eta}^{\dot\alpha\, j}_{ B} 
\left( F^{\beta\gamma\, i}_k\stackrel{\leftrightarrow}{\mathcal D}_{\alpha\dot\alpha} F_{\beta\gamma\, j}^k\right)
\label{8}
\eeq
minus trace in $A,B$ (the gluon field strength tensors are given above in the spinorial notation).
It is easy to see, however, that even if $\langle {\mathcal O}^A_B\rangle \neq 0$,
the chiral SU(8) is broken not completely, but rather down to U(1)$^7$ at best. (In fact, we would have 
 U(1)$^8$, see below). This is unsatisfactory 
 since in this case we will have to match the residual 't Hooft triangles, which does not seem possible.
 
 The following operator built of six fermion fields 
 \beq
 {\mathcal O}^{ABA^\prime B^\prime} = \left(\eta^{\alpha\, A}_i\,\chi^{\{ij\}}_\beta \, \eta_{\alpha\, j}^B\right)
 \left(\eta^{\alpha^\prime \, A^\prime}_{i^\prime}\,\chi^{\{\beta\, i^\prime j^\prime\}}\, \eta_{\alpha^\prime\, j^\prime}^{B^\prime}\right)
 \label{9}
 \eeq
  is
 the lowest-dimension operator breaking the global symmetry in the $\eta$ sector completely.
Despite its rather contrived structure, a non-vanishing  expectation 
value $\langle  {\mathcal O}^{ABA^\prime B^\prime}\rangle$  is not ruled out apriori.
Therefore, it is natural to
assume that U(8) is spontaneously broken. Then  64
 Goldstone bosons (``pions") appear. The vacua can no longer be discrete, 
 since the presence of pions means that the vacuum manifolds are continuous (albeit compact).
Instead of having a set of discrete vacuum points, we have a continuous extension around each point. 
We will return to discussion of this aspect of the hybrid theory later.

A few words about  the extra U(1) symmetry showing up upon inclusion of the $\eta$ fields.
First, the {\em conserved} current in (\ref{6}) -- the one that is analogous to the vector current and does not belong to
the common sector --
now takes the form
\beq 
\tilde{j}^{\dot\alpha\alpha}_{(1)}
=
j^{\dot\alpha\alpha}_{(\psi )} - 
j^{\dot\alpha\alpha}_{(\chi )}
+\frac{1}{2} \sum_{A=1}^8 \bar\eta^{\dot\alpha} \eta^\alpha
\label{10}
\eeq
Note that the operator (\ref{9}) is invariant under transformations generated
by the current $\tilde{j}^{\dot\alpha\alpha}_{(1)}$. Hence, its vacuum expectation value does not break
the corresponding vector-like symmetry. This is a remarkable circumstance.

In addition, one can
consider the following currents:
\beq
\tilde{j}^{\dot\alpha\alpha}_{(2)}= j^{\dot\alpha\alpha}_{(\psi )} + 
j^{\dot\alpha\alpha}_{(\chi )} - \frac{N}{4} \sum_{A=1}^8 \bar\eta^{\dot\alpha} \eta^\alpha\,,\qquad j^{\dot\alpha\alpha}_{(3)} = \sum_{A=1}^8 \bar\eta^{\dot\alpha} \eta^\alpha\,.
\label{11}
\eeq
Unlike $j^{\dot\alpha\alpha}_{(2)}$, the   current $\tilde{j}^{\dot\alpha\alpha}_{(2)}$
is anomaly-free, 
while the last one is anomalous. Accounting for $\tilde{j}^{\dot\alpha\alpha}_{(2)}$, we extend the
SU(8) global symmetry of the $\eta$ sector to U(8).
The remnant of the anomalous $j^{\dot\alpha\alpha}_{(3)}$ is a discrete 
$Z_8$ symmetry, which is not broken by the condensate (\ref{8}). It is broken down to $Z_4$ by the condensate 
(\ref{9}).

The presence of the massless pions, even though they are not in the common sector, somewhat dilutes
the concept of  
planar equivalence between our hybrid theory and supersymmetric gluodynamics. Indeed, the latter theory, having $N$ 
{\em discrete} vacua, supports a number of BPS-saturated domain walls, whose tension is determined
by the difference of the gluino condensates
in the vacua between which the given wall interpolates \cite{DS}.
In the hybrid theory the vacuum manifold is continuous. Under these circumstances, strictly speaking, there are no
domain walls. More exactly, the would-be walls will have a double-layer structure: a finite-thickness core, and infinite-thickness pion tails attached to it. Although the pion tails are suppressed by $1/N$, their contribution to the
tension is actually infinite, no matter how large $N$ is. This seems to correlate with the fact that the operator $\lambda^2$ has no projection onto the hybrid theory.

\vspace{5mm}
We are  very grateful to Mithat \"Unsal for valuable discussions. This work is supported in part by DOE grant DE-FG02- 94ER-40823.

\end{document}